\documentclass[11pt]{article}

\usepackage[T1]{fontenc}
\usepackage[utf8]{inputenc}
\usepackage[english]{babel}
\usepackage{algorithm}
\usepackage{algpseudocode}
\usepackage{doi}
\usepackage{booktabs}
\usepackage{longtable}
\usepackage{amssymb} 
\usepackage{amsthm}

\newtheorem{definition}{Definition}

\usepackage{tikz}
\usetikzlibrary{arrows.meta, positioning, shapes.geometric}

\usepackage[a4paper,margin=30mm]{geometry}
\usepackage{lmodern}
\usepackage{microtype}
\usepackage{setspace}
\usepackage{amsmath,amssymb}

\usepackage{booktabs}
\usepackage{tabularx}

\usepackage[numbers]{natbib}
\usepackage{placeins} 

\title{Geopolitical and Institutional Constraints on Adaptive Market Efficiency\\
	\large A Feasibility Diagnostic for Robust Portfolio Construction}

\author{
	Roberto Garrone\\
	\vspace{0.5em}
	\small Preprint
}

\author{
	Roberto Garrone\\[0.5ex]
	\small
	University of Salford, Salford, UK\\
		\small
	University Sapienza Unitelma, Rome, Italy\\
		\small
	University of Milan-Bicocca, Milan, Italy\\
		\small
	\texttt{roberto.garrone@unimib.it}
}

\date{\today}

\begin{document}
	\maketitle
	
\begin{abstract}
This paper develops a structural framework for characterizing the informational feasibility of financial markets under heterogeneous institutional and geopolitical conditions. Departing from the assumption of uniform and time-invariant market efficiency, we conceptualize adaptive efficiency as a localized and state-dependent property emerging from the interaction between economic scale, institutional enforcement, and geopolitical embedding. To operationalize this perspective, we introduce the Geopolitical–Adaptive Efficiency Ratio (GAER), a descriptive cross-sectional indicator measuring the concentration of adaptive-efficiency-supporting mass within institutionally and geopolitically central assets. GAER is not a return-predictive signal, factor, or regime classifier. Instead, it functions as a diagnostic boundary condition, delimiting the domain in which ranking-based, robustness-oriented portfolio construction methods are plausibly applicable.

The framework integrates insights from adaptive market theory, institutional economics, and political economy, linking disclosure continuity, liquidity provision, and enforcement credibility to the persistence of informational signals in asset prices. It is intended for portfolio researchers and practitioners employing robustness-oriented and constraint-aware methods rather than return-forecasting models. We formalize GAER, discuss its theoretical properties, and illustrate its interpretation using verosimilar global equity data. Applications include universe conditioning, cross-market comparability, stress-period diagnostics, and feasibility checks for factor-based strategies. By separating where adaptive efficiency plausibly holds from how portfolios should be constructed, this paper provides a conceptual foundation for constraint-aware financial modeling and complements robustness-oriented portfolio construction frameworks without relying on forecast-driven assumptions or parametric optimization.
\end{abstract}

\noindent\textbf{Keywords:}
adaptive market efficiency; institutional quality; geopolitical risk;
informational feasibility; portfolio construction; robustness;
market structure; liquidity; non-predictive diagnostics.

\section{Introduction: Localizing Adaptive Market Efficiency}
\label{sec:intro}

The Efficient Market Hypothesis (EMH) is commonly interpreted as implying that prices
incorporate available information rapidly and reliably across the investable universe
\citep{Fama1970,Fama1991}. Yet a long empirical tradition documents persistent deviations,
slow adjustment, and instability in the mapping from information to prices
\citep{Shiller1981,Stiglitz1980}. The Adaptive Markets Hypothesis (AMH) offers a
reconciliation: market efficiency is not a universal primitive but a conditional and
evolving property shaped by competition, learning, and market ecology
\citep{Lo2004,Lo2017}. Building on this view, the present preprint adopts an adaptive and
institutionally grounded interpretation of efficiency in which informational performance is
\emph{local, conditional, and state-dependent}, emerging endogenously from the interaction
between institutional capacity, liquidity conditions, regulatory enforcement, and
geopolitical stability \citep{North1990,AcemogluRobinson2012,Williamson1985}.

\paragraph{From universal postulate to feasibility question.}
Under a constraint-aware perspective, efficiency is not assumed to hold uniformly; rather,
it is treated as an \emph{emergent property of the environment} in which information is
generated, transmitted, and enforced. Informational signals are incorporated more rapidly
and reliably where market depth is high, disclosure standards are enforced, and credible
backstops reduce tail-event discontinuities \citep{Chordia2008}. Conversely, in segments
characterized by low liquidity, weaker enforcement, or institutional fragility,
informational frictions persist and price discovery becomes noisier or episodic
\citep{Stiglitz1980,Shiller1981}. This implies that the effective domain over which
adaptive efficiency holds is constrained by institutional and geopolitical conditions,
not determined solely by asset-level characteristics.

\paragraph{Geopolitics as infrastructure for informational continuity.}
Institutional heterogeneity extends naturally to geopolitics. Monetary dominance,
sanctions regimes, trade fragmentation, and crisis-time interventions shape the stability
of information environments by reallocating liquidity and enforcement capacity across
jurisdictions and firms. Historically, systemic stabilizers have been unevenly applied,
with core currency areas and institutionally central markets more likely to receive
support during stress \citep{Kindleberger1973,Eichengreen2011,Shin2010}. This asymmetry
implies that currency regimes supported by fiscal and military capacity, and globally
integrated firms operating within stable legal frameworks, tend to exhibit greater
continuity of disclosure and price discovery than markets operating outside these
structures \citep{Eichengreen2011,Shin2010}. From this perspective, informational
efficiency is not uniform but conditioned by the stability of the surrounding
institutional and geopolitical infrastructure.

\paragraph{Firm-level implication: systemic embedding and signal persistence.}
Under strong-form EMH, firm size and geopolitical embeddedness should be largely
irrelevant for informational efficiency once risk is correctly priced. In contrast, an
adaptive and institutionally grounded view predicts that large, systemically important
firms operating within stable regulatory and monetary environments benefit from more
continuous disclosure, intensive monitoring, and privileged access to liquidity,
reducing information disruption and improving signal persistence
\citep{Coase1937,Williamson1985,Shin2010}. Firms lacking such embedding are more exposed to
funding constraints, regulatory discontinuities, and abrupt repricing, particularly under
geopolitical or macroeconomic stress \citep{Chordia2008,Shiller1981}. What appear as
``anomalies'' in frictionless EMH frameworks can therefore be interpreted as predictable
outcomes of heterogeneous information environments rather than transient inefficiencies.

\paragraph{Portfolio-relevant implication (without tactical claims).}
These considerations matter for portfolio research because many robustness-oriented
constructions rely on relative rankings, bounded adjustments, and constrained turnover.
Such methods implicitly require a minimum degree of information stability that is not
uniformly available across assets and jurisdictions \citep{DeMiguel2009,KoijenRossi2019}.
Restricting attention to assets embedded in stable institutional and geopolitical
environments is therefore not a rejection of efficiency, but a recognition that
efficiency is unevenly distributed and must be localized. The present paper addresses
this localization problem by introducing a descriptive diagnostic that summarizes the
concentration of adaptive-efficiency-supporting conditions in the investable universe.

\paragraph{Scope and intended contribution.}
This paper contributes to the portfolio construction and asset management literature by
introducing a diagnostic framework for assessing the informational feasibility of
ranking-based and robustness-oriented methods. It does not propose a new asset pricing
model, return factor, or equilibrium theory. Instead, it focuses on characterizing the
environmental conditions under which existing portfolio methodologies can be expected to
operate with stable informational inputs.

\subsection{Research Gap and Contribution}
\label{sec:research_gap}

Despite extensive advances in asset pricing, portfolio construction, and adaptive
market theory, the literature lacks a formal framework for characterizing the
\emph{informational feasibility domain} within which robustness-oriented portfolio
methods are plausibly applicable. Existing strands of research address related
components in isolation, but do not integrate them into a coherent diagnostic layer
situated between theory and implementation.

First, classical and modern formulations of market efficiency---from the EMH to the
AMH---conceptualize efficiency as either universal in principle or evolving through
competition and learning \citep{Fama1970,Fama1991,Lo2004,Lo2017}. While these frameworks
acknowledge time variation and adaptation, they remain largely silent on how the
\emph{continuity of informational processing} varies structurally across assets and
jurisdictions. Empirically observed deviations are typically attributed ex post to
frictions, behavioral effects, or limits to arbitrage \citep{Stiglitz1980,Shiller1981},
rather than treated ex ante as consequences of heterogeneous institutional and
geopolitical environments.

Second, the portfolio construction and asset management literature has developed a wide
range of robustness-oriented methods---including equal-weighting, rank-based allocations,
bounded adjustments, and constraint-aware optimizations---explicitly motivated by
estimation error, non-stationarity, and model uncertainty
\citep{fernholz2002,Pedersen2015}. However, these approaches generally \emph{take the
	informational domain as given}, relying on liquidity or coverage screens without a
principled characterization of whether the underlying environment supports sufficiently
stable signal extraction in the first place. As a result, portfolio robustness is often
addressed at the level of allocation mechanics rather than informational feasibility.

Third, a substantial literature focuses on liquidity, trading frictions, and
implementation costs, providing sophisticated diagnostics for \emph{economic feasibility},
such as illiquidity measures, turnover penalties, and transaction-cost models
\citep{Amihud2002,pastor2003}. While these contributions are central for
assessing realized performance, they address how portfolios can be implemented
profitably, not where adaptive efficiency plausibly holds. Informational feasibility and
economic feasibility are therefore often conflated or treated sequentially without an
explicit conceptual boundary between them.

Finally, insights from institutional economics and political economy emphasize the role
of enforcement quality, governance, and geopolitical stability in shaping market outcomes
\citep{North1990,AcemogluRobinson2012,Williamson1985,Kindleberger1973,Eichengreen2011}.Yet these insights have rarely been operationalized into \emph{model-agnostic diagnostics}
usable by portfolio researchers without turning geopolitical considerations into
return-predictive signals or discretionary overlays.

This paper addresses this gap by introducing the \emph{Geopolitical--Adaptive Efficiency Ratio (GAER)}, a descriptive, cross-sectional diagnostic that summarizes the concentration of adaptive-efficiency-supporting conditions within an investable universe.
By design, GAER does not forecast returns, define regimes, or prescribe portfolio weights.
Instead, it formalizes a missing intermediate layer between theory and application: a
feasibility boundary that clarifies \emph{where} ranking-based and robustness-oriented
portfolio methods can be expected to operate under relatively continuous informational
conditions. In doing so, the framework complements existing portfolio construction and
execution methodologies while preserving a clear separation between informational
feasibility, economic extraction, and allocation design
\citep{HarveyLiuZhu2016,KoijenRossi2019}.

\subsection{Theoretical Grounding}
\label{subsec:theory_grounding}

The interpretation above aligns with established results in institutional and monetary
economics emphasizing that markets require enforceable contracts, credible governance,
and institutional continuity to function effectively
\citep{Smith1776,Ricardo1817,Mill1848}. Subsequent contributions highlighted the role of
organizational scale and structure in shaping market outcomes \citep{Marx1867,Marshall1890}.
Keynes emphasized the role of conventions, policy support, and liquidity provision in
sustaining orderly markets under uncertainty \citep{Keynes1936}, while Hayek showed that
price systems aggregate dispersed information efficiently only when legal and institutional
filters preserve signal integrity \citep{Hayek1945}. Coase provided a micro-level foundation
by demonstrating that economic activity concentrates when internal coordination reduces
transaction costs, particularly under regulatory or political fragmentation \citep{Coase1937},
a mechanism reinforced by public choice insights on endogenous regulation and differential
enforcement \citep{BuchananTullock1962}. Modern institutional economics links enforcement
quality and political stability to long-run economic performance \citep{AcemogluRobinson2012}.

\section{Adaptive Efficiency as an Emergent Property}
\label{sec:adaptive_efficiency}

Under strong-form EMH, informational efficiency is often treated as homogeneous in
principle, with deviations attributed to frictions, limits to arbitrage, or behavioral
effects \citep{Fama1970,Fama1991}. Under AMH, efficiency is instead local and
state-dependent, shaped by competition, learning, and market ecology
\citep{Lo2004,Lo2017}. A central implication for applied modeling is that the stability
of informational signals is contingent on the institutional and infrastructural
environment that produces and enforces those signals.

Institutional economics provides a structural basis for this contingency. Markets require
enforceable contracts, credible governance, and continuity of rules to support reliable
information aggregation and exchange \citep{North1990,AcemogluRobinson2012,Williamson1985}.
When enforcement weakens or becomes discontinuous, informational frictions can persist
even in liquid markets, and the mapping from fundamentals and news to prices becomes
noisier or episodic \citep{Stiglitz1980,Chordia2008}.

\paragraph{Interpretation.}
In this view, ``efficiency'' should be treated as an emergent property of an environment
rather than a binary label for an asset. This motivates a diagnostic that summarizes
where adaptive-efficiency-supporting conditions are concentrated.

\section{Institutional and Geopolitical Determinants of Informational Stability}
\label{sec:inst_geo_embedding}

Beyond domestic institutional quality, geopolitics shapes the information environment
through monetary dominance, sanctions regimes, trade fragmentation, and crisis-time
interventions. Historically, systemic stabilizers and liquidity backstops have been
unevenly applied, with central currency areas and institutionally core markets more
likely to receive supportive interventions during stress
\citep{Kindleberger1973,Eichengreen2011,Shin2010}.

At the firm level, large and systemically important issuers embedded in stable legal and
monetary environments may benefit from continuous disclosure, intensive monitoring,
privileged access to liquidity, and implicit backstops---all of which reduce information
disruption and support continuity in price discovery \citep{Williamson1985,Shin2010}.
By contrast, assets operating in peripheral or institutionally fragile contexts may
exhibit higher discontinuity in disclosure, enforcement, and market-making capacity.

\paragraph{Implication.}
If informational continuity is heterogeneous, then the effective domain over which
ranking-based methods (and other robustness-oriented approaches) can be expected to
behave predictably is also heterogeneous.

\section{From Theory to Measurement: Defining GAER}
\label{sec:gaer_definition}

\begin{definition}[Geopolitical--Adaptive Efficiency Ratio (GAER)]
	\label{def:gaer}
	Let $\mathcal{U}_t$ denote the investable universe at time $t$, and let
	$\mathcal{U}_t^{\mathrm{core}} \subseteq \mathcal{U}_t$ denote a subset classified as
	``systemically embedded'' (e.g., institutionally central and/or mega-cap assets). For each
	asset $i$, let $\mathrm{MC}_{i,t}$ denote economic scale (e.g., market capitalization),
	$G_{i,t}\in[0,1]$ a normalized indicator of geopolitical embedding, and
	$I_{i,t}\in[0,1]$ a normalized proxy for institutional quality relevant to information
	processing. Define:
	\begin{equation}
		\label{eq:gaer}
		\mathrm{GAER}_t \;=\;
		\frac{\sum_{i \in \mathcal{U}^{\mathrm{core}}_t} \mathrm{MC}_{i,t}\,G_{i,t}\,I_{i,t}}
		{\sum_{j \in \mathcal{U}_t} \mathrm{MC}_{j,t}\,G_{j,t}\,I_{j,t}}.
	\end{equation}
\end{definition}

The classification of the core subset $\mathcal{U}_t^{\mathrm{core}}$ is intentionally
left flexible and may reflect economic scale thresholds, institutional centrality,
or regulatory relevance, depending on the application.

By construction, $\mathrm{GAER}_t\in[0,1]$. It can be interpreted as the share of aggregate
\emph{adaptive-efficiency-supporting mass} concentrated in the core segment of the
universe. Higher values indicate that a larger fraction of economic scale is embedded in
environments where information flows, enforcement, and liquidity provision are more
continuous and reliable.

\paragraph{State variable, not signal.}
GAER is a slow-moving diagnostic summarizing structural conditions. It is not designed
for short-horizon inference, return prediction, portfolio weighting, market timing, or
dynamic exposure management. This positioning is important in light of concerns about
factor proliferation and ex post narrative fitting \citep{HarveyLiuZhu2016}.

\paragraph{Complementarity with execution and implementation feasibility.}
GAER concerns \emph{informational feasibility}: whether adaptive efficiency is plausibly
persistent enough for ranking-based, constraint-aware constructions to be meaningful. It
does not address \emph{economic feasibility}: whether any residual structure can be
captured after turnover, liquidity, and capital frictions. Those questions belong to
execution and implementation diagnostics (e.g., liquidity risk measures and trading-cost
frameworks) and should remain analytically distinct \citep{Amihud2002,pastor2003}.

\section{Interpretation and Theoretical Properties}
\label{sec:gaer_properties}

\subsection{Non-claims and boundary role}
GAER does not test strong-form efficiency, nor does it imply that assets outside
high-GAER environments are inefficient, mispriced, or dominated. Its interpretation is
strictly comparative and contextual: GAER indicates where adaptive-efficiency-supporting
conditions are more likely to persist with sufficient continuity to justify the use of
ranking-based and robustness-oriented methods \citep{Lo2004,Pedersen2015}.

\subsection{Basic properties}
Let $w_{i,t} := \mathrm{MC}_{i,t}G_{i,t}I_{i,t} \ge 0$. Then GAER is a concentration ratio:
\[
\mathrm{GAER}_t \;=\; \frac{\sum_{i \in \mathcal{U}_t^{\mathrm{core}}} w_{i,t}}
{\sum_{j \in \mathcal{U}_t} w_{j,t}}.
\]
Properties follow immediately:
\begin{itemize}
	\item \textbf{Scale--embedding interaction.} High $\mathrm{MC}$ alone does not dominate if $G$ or $I$ is low.
	\item \textbf{Comparability.} For a fixed construction of $G$ and $I$, GAER supports cross-market comparison as a summary statistic.
	\item \textbf{Stress sensitivity.} In stress episodes, shifts in $\mathrm{MC}$ (relative repricing) and/or in embedding indicators may alter GAER,
	contextualizing changes in feasibility.
\end{itemize}

\subsection{Why a diagnostic is useful even without forecasting}
A large portion of robustness-oriented practice relies on cross-sectional rankings and
bounded adjustments because these are comparatively stable under estimation error
\citep{Pedersen2015,fernholz2002}. However, even rank stability depends on the continuity
of information processing and enforcement. GAER provides a compact descriptor of that
continuity domain, without converting geopolitics into an investable signal.

\section{Illustrative Global Snapshot}
\label{sec:illustration}

This section illustrates the interpretation of GAER using a global equity
snapshot in which the core set is defined as the top cumulative market-capitalization
subset (e.g., 75\%), and the numerator aggregates $\mathrm{MC}\cdot G\cdot I$ over that
set. The purpose is interpretive rather than empirical: the values reported below are
illustrative and constructed using publicly available information and qualitative approximations, with the sole aim of clarifying the mechanics and interpretation of GAER, not of providing calibrated estimates or empirical inference.

\begin{longtable}{llrrrrrc}
	\caption{GAER input snapshot (illustrative values; MCAP and ADV in USD billions). Core denotes the top 75\% cumulative market-capitalization subset used for the numerator of GAER.}\label{tab:gaer_hypothetical}\\
	\toprule
	Ticker & Reg. & \multicolumn{1}{r}{MCAP} & \multicolumn{1}{r}{ADV} & \multicolumn{1}{r}{$G$} & \multicolumn{1}{r}{$I$} & \multicolumn{1}{r}{$w=\mathrm{MCAP}\cdot G\cdot I$} & Core\\
	\midrule
	\endfirsthead
	
	\toprule
	Ticker & Reg. & \multicolumn{1}{r}{MCAP} & \multicolumn{1}{r}{ADV} & \multicolumn{1}{r}{$G$} & \multicolumn{1}{r}{$I$} & \multicolumn{1}{r}{$w$} & Core\\
	\midrule
	\endhead
	
	\midrule
	\multicolumn{8}{r}{\footnotesize Continued on next page.}\\
	\endfoot
	
	\bottomrule
	\endlastfoot
	
	MSFT & US & 3,100 & 14.0 & 0.95 & 0.96 & 2,827.2 & \checkmark \\
	AAPL & US & 2,900 & 18.0 & 0.95 & 0.95 & 2,617.2 & \checkmark \\
	AMZN & US & 1,900 & 10.0 & 0.94 & 0.94 & 1,678.8 & \checkmark \\
	NVDA & US & 2,000 & 22.0 & 0.92 & 0.90 & 1,656.0 & \checkmark \\
	GOOGL & US & 1,800 & 9.0 & 0.93 & 0.93 & 1,556.8 & \checkmark \\
	META & US & 1,200 & 8.0 & 0.92 & 0.92 & 1,015.7 & \checkmark \\
	BRK.B & US & 900 & 1.5 & 0.95 & 0.90 & 769.5 & \checkmark \\
	2222.SR & SA & 2,000 & 0.6 & 0.60 & 0.68 & 816.0 & \checkmark \\
	TSLA & US & 800 & 12.0 & 0.88 & 0.85 & 598.4 & \checkmark \\
	LLY & US & 750 & 3.5 & 0.93 & 0.88 & 613.8 & \checkmark \\
	AVGO & US & 700 & 4.0 & 0.93 & 0.90 & 585.9 & \checkmark \\
	V & US & 520 & 2.8 & 0.92 & 0.90 & 430.6 & \checkmark \\
	JPM & US & 500 & 3.0 & 0.90 & 0.88 & 396.0 & \checkmark \\
	MA & US & 470 & 2.2 & 0.92 & 0.90 & 389.2 & \checkmark \\
	UNH & US & 450 & 1.6 & 0.90 & 0.87 & 352.4 & \checkmark \\
	XOM & US & 450 & 2.5 & 0.89 & 0.86 & 344.4 & \checkmark \\
	JNJ & US & 430 & 2.0 & 0.91 & 0.89 & 348.3 & \checkmark \\
	WMT & US & 420 & 2.2 & 0.91 & 0.88 & 335.0 & \checkmark \\
	TSM & TW & 600 & 4.5 & 0.75 & 0.82 & 369.0 & \checkmark \\
	ASML & NL & 420 & 1.2 & 0.85 & 0.88 & 314.2 &  \\
	PG & US & 380 & 2.0 & 0.91 & 0.88 & 304.3 &  \\
	HD & US & 380 & 1.7 & 0.90 & 0.88 & 300.7 &  \\
	COST & US & 360 & 1.8 & 0.91 & 0.88 & 287.7 &  \\
	ORCL & US & 360 & 2.4 & 0.90 & 0.87 & 281.9 &  \\
	NESN.SW & CH & 330 & 0.4 & 0.86 & 0.86 & 244.3 &  \\
	005930.KS & KR & 380 & 1.8 & 0.78 & 0.80 & 237.1 &  \\
	BAC & US & 310 & 2.2 & 0.88 & 0.86 & 234.8 &  \\
	KO & US & 270 & 1.6 & 0.90 & 0.87 & 211.4 &  \\
	PEP & US & 260 & 1.5 & 0.90 & 0.87 & 203.6 &  \\
	NFLX & US & 260 & 5.0 & 0.90 & 0.86 & 201.2 &  \\
	TM & JP & 270 & 0.9 & 0.82 & 0.84 & 186.0 &  \\
	ROG.SW & CH & 240 & 0.4 & 0.86 & 0.85 & 175.4 &  \\
	700.HK & HK & 320 & 1.0 & 0.70 & 0.75 & 168.0 &  \\
	NOVN.SW & CH & 220 & 0.5 & 0.86 & 0.85 & 160.8 &  \\
	OR.PA & FR & 230 & 0.5 & 0.83 & 0.84 & 160.4 &  \\
	AZN & UK & 210 & 0.8 & 0.85 & 0.85 & 151.7 &  \\
	SAP & DE & 210 & 0.7 & 0.84 & 0.85 & 150.2 &  \\
	SHEL & UK & 200 & 0.9 & 0.84 & 0.84 & 141.1 &  \\
	MC.PA & FR & 200 & 0.6 & 0.83 & 0.84 & 139.4 &  \\
	SIE.DE & DE & 170 & 0.5 & 0.84 & 0.84 & 119.9 &  \\
	HSBA & UK & 150 & 0.8 & 0.84 & 0.83 & 104.6 &  \\
	RELIANCE.NS & IN & 230 & 0.7 & 0.65 & 0.70 & 104.7 &  \\
	TTE & FR & 150 & 0.7 & 0.82 & 0.82 & 100.9 &  \\
	SONY & JP & 150 & 0.7 & 0.80 & 0.83 & 99.6 &  \\
	9983.T & JP & 150 & 0.9 & 0.80 & 0.80 & 96.0 &  \\
	TCS.NS & IN & 170 & 0.4 & 0.66 & 0.72 & 80.8 &  \\
	TCEHY & CN & 400 & 2.5 & 0.55 & 0.72 & 158.4 &  \\
	BABA & CN & 180 & 3.0 & 0.55 & 0.70 & 69.3 &  \\
	BIDU & CN & 100 & 1.2 & 0.55 & 0.68 & 37.4 &  \\
	\footnote{
		GAER inputs can be sourced from standard market data providers (market capitalization,
		trading volume), international financial institutions (IMF, BIS), governance databases
		(World Bank WGI, OECD), and regulatory bodies (IOSCO, WFE). The table reports illustrative values constructed from publicly available information and qualitative normalization,
		not calibrated estimates.
	}
\end{longtable}

Table~\ref{tab:gaer_hypothetical} reports a GAER snapshot for a globally
diversified large-cap equity universe. The core subset—defined as the top 75\% of
cumulative market capitalization—accounts for a disproportionately large share of the
weighted mass \( w=\mathrm{MCAP}\cdot G \cdot I \), resulting in a relatively high GAER
value (approximately 0.77). This concentration reflects the systematic co-location of
economic scale, liquidity, and institutional and geopolitical embedding within a limited
set of mega-capitalization firms, predominantly operating in dominant currency areas and
subject to strong disclosure standards, regulatory oversight, and market infrastructure.
Assets outside the core generally exhibit lower values of \(G\) and/or \(I\)—often
associated with weaker geopolitical integration indicators, thinner information
environments, or less continuous enforcement—which reduces their contribution to
aggregate adaptive efficiency despite, in some cases, substantial market capitalization.
More broadly, the table illustrates that in globally diversified large-cap universes,
adaptive-efficiency-supporting mass tends to concentrate in institutionally central
issuers, leading to higher GAER values when core markets dominate price discovery
capacity, and lower values when a larger fraction of economic scale resides in weaker or
more discontinuous informational environments. In this sense, GAER summarizes the
structural concentration of conditions supportive of adaptive efficiency and provides a
transparent motivation for universe conditioning in robustness-oriented portfolio
construction, without implying return predictability or tactical allocation guidance.

\section{Applications as Feasibility Diagnostics}
\label{sec:applications}

\subsection{Universe conditioning}
GAER can delimit the investable universe for portfolio construction methods that rely on
relative rankings, bounded adjustments, or turnover control. When GAER is high,
restricting attention to the core subset may preserve informational continuity assumptions
without materially reducing adaptive-efficiency-supporting mass. When GAER is low, the
informational domain may violate assumptions required for reliable ranking and bounded
optimization, analogous to liquidity and coverage screens.

\subsection{Cross-market comparability}
GAER supports formal comparison of informational environments across equity universes
(e.g., U.S.-centric, developed ex-U.S., emerging markets). Differences in GAER can be
interpreted as reflecting variation in institutional depth, regulatory continuity, and
geopolitical embedding, without invoking mispricing claims.

\subsection{Stress-period diagnostics}
During macroeconomic or geopolitical stress, liquidity reallocation and repricing may
shift GAER. While GAER is not a rebalancing trigger, it can contextualize observed
changes in turnover, tracking error, or factor instability by documenting shifts in the
concentration of adaptive efficiency across the universe.

\subsection{Feasibility checks for factor-based modeling}
Factor models and cross-sectional ranking methods can be supplemented with GAER as a
feasibility covariate. Low GAER does not imply factor failure, but signals that the
information environment may be discontinuous relative to the assumptions required for
stable ranking and bounded construction \citep{HarveyLiuZhu2016,KoijenRossi2019}.

\section{Implications for Portfolio Construction}
\label{sec:relation_portfolios}

GAER addresses the question: \emph{where is adaptive efficiency plausibly continuous
	enough to support robustness-oriented methods?} It does not prescribe portfolio weights
or define an allocation algorithm. Downstream portfolio construction can treat GAER as an
exogenous feasibility boundary and remain fully implementable without computing GAER in
real time.

\paragraph{Complementarity with friction-focused feasibility ratios.}
Whereas GAER concerns informational feasibility (continuity of information processing),
implementation feasibility depends on trading frictions and capital constraints (e.g.,
liquidity costs and turnover). This paper deliberately separates the informational domain
question from economic extraction questions, which are best handled by execution and
implementation diagnostics \citep{Amihud2002,pastor2003}.

\section{Limitations and Open Directions}
\label{sec:limitations}

The construction of GAER depends on the specification of the geopolitical embedding
indicator $G$ and the institutional quality proxy $I$, which necessarily involves
measurement choices and potential sensitivity to proxy selection. As a result, GAER
should be interpreted as a diagnostic framework rather than as a uniquely defined
index.

Several directions for future research follow naturally from this formulation:
\begin{itemize}
	\item systematic design and validation of proxies for geopolitical embedding and
	institutional quality;
	\item robustness analysis with respect to alternative normalizations, weighting
	schemes, and definitions of the core asset subset;
	\item construction of historical GAER time series to examine structural shifts in
	the concentration of adaptive-efficiency-supporting conditions across market
	regimes and crisis episodes.
\end{itemize}

The present preprint deliberately emphasizes conceptual formalization and interpretive
clarity over empirical calibration. Extensions involving calibrated implementations,
sensitivity analysis across alternative proxy choices, or explicitly time-varying
applications are left for future work. This separation is intentional: it preserves
the diagnostic and non-predictive nature of GAER and avoids conflating informational
feasibility with economic performance or portfolio outcomes.

Accordingly, the objective of this contribution is not to propose a definitive or
exhaustive index, but to introduce a reusable \emph{measurement template} for
feasibility-aware financial modeling that can be adapted and refined as empirical
and institutional data evolve.

\section{Conclusion}
\label{sec:conclusion}

This paper argues that adaptive market efficiency is not uniformly distributed across
assets and jurisdictions because the continuity of disclosure, enforcement, and liquidity
provision is heterogeneous. GAER operationalizes this heterogeneity as a descriptive
concentration ratio, summarizing where adaptive-efficiency-supporting mass is embedded.
Properly interpreted, GAER is a diagnostic boundary condition---not a return signal---that
complements robustness-oriented and constraint-aware financial modeling. By explicitly separating informational feasibility from economic extraction and
portfolio optimization, the GAER framework provides a transparent bridge between
institutional context and applied portfolio methodology, without introducing
forecasting assumptions or discretionary geopolitical judgments.

	\section*{Reproducibility}
	
	All components of the methodology are algorithmically defined. A reference implementation requires only total-return prices, basic liquidity measures, and optional fundamental data.


\end{document}